\documentclass[runningheads]{llncs}
\usepackage[T1]{fontenc}
\usepackage{graphicx}
\usepackage{multicol}
\begin{document}
\title{Coalescing MPI communication in 6D Vlasov simulations: solving ghost domains in Vlasiator}
\titlerunning{Coalescing MPI in 6D: solving ghost domains in Vlasiator}
%
\author{Markus Battarbee\inst{1}\orcidID{0000-0001-7055-551X} \and
Urs Ganse\inst{1}\orcidID{0000-0003-0872-1761} \and
Yann Pfau-Kempf\inst{1}\orcidID{0000-0001-5793-7070} \and
Markku Alho\inst{1}\orcidID{0000-0001-9762-6795} \and 
Konstantinos Papadakis\inst{1}\orcidID{0000-0002-3307-6015} \and 
Minna Palmroth\inst{1,2}\orcidID{0000-0003-4857-1227}}%
\authorrunning{M. Battarbee et al.}
%
\institute{Department of Physics, University of Helsinki, Helsinki, Finland
\email{markus.battarbee@helsinki.fi}
\and
Finnish Meteorological Institute, Helsinki, Finland\\
}
\maketitle              
\begin{abstract}
High-performance computing is used for diverse simulations, some of which parallelize over the Message Passing Interface (MPI) with ease, whilst others may have challenges related to uniform balancing of computational load and communication between simulation domains. We introduce an alternative approach to solving advection equations, specifically in an application to solving the six-dimensional Vlasov equation for modelling space plasmas. Communicating larger ghost domains around the partition assigned to each MPI task and computing on these ghost cells allows for coalescing several discrete communication calls into one. This approach needs more overall data communication and computation, but provides interesting new avenues for the balancing of computational load between MPI tasks. We discuss this trade-off, how it may assist in developing other algorithmic improvements, and how the transition to heterogeneous CPU-GPU architectures may impact its usefulness.

\end{abstract}
\section{Introduction}

In the quest for greater understanding of physical phenomena, computational scientists are designing ever more accurate and larger simulations. Lately, increase in computing power has transitioned from increasing clock speeds to enabling greater parallelism. Modern high-performance computing (HPC) clusters may offer thousands of compute nodes, each powering several central processing units (CPUs) and/or graphics processing units (GPUs), each containing further parallelism through, for example, compute cores and streaming multiprocessors.

This foundational transformation in the availability of processing power brings significant challenges to the development and optimization of many scientific
applications. Unless the computational task is \emph{ridiculously parallelizeable} (where the result of one computational domain has no dependence or impact on other domains), the efficiency of the task is impacted by the communication interfaces, such as the Message Passing Interface (MPI) software and the hardware interconnects between computation nodes. Additionally, the design of the simulation parallelism and the suitability of the algorithms for said purpose may become critically important. 

In this work, we discuss load distribution and parallel efficiency of Vlasiator \cite{Palmroth2018LRCA}, a space plasma simulation code which solves the Vlasov equation for kinetic ions. We explain challenges related to the dynamically evolving computational complexity of the Vlasiator grid, as well as how adaptive mesh refinement (AMR) influences parallel efficiency.

\section{Vlasiator}

Vlasiator \cite{Palmroth2018LRCA,vonAlfthan2014JASTP} is a HPC hybrid-Vlasov space plasma simulation code which models plasma dynamics by propagating kinetic ions (usually single-species protons) on a 6-dimensional (6D) Cartesian mesh. Unlike particle-in-cell \cite{pic} codes, Vlasiator does not propagate macroparticles, instead tracking the phase-space density $f_s(\textbf x,\textbf v)$ of charged species $s$ directly with a grid-discretized distribution function representation. Vlasiator is mainly tailored towards large-scale simulations of the terrestrial magnetosphere and its surrounding magnetosheath--bow shock--foreshock system. With a focus on macro-to-ion-scale dynamics, electrons are considered a charge-neutralizing fluid. Electromagnetic fields $(\vec{E},\vec{B})$ are solved using Ohm's law from magneto-hydrodynamics including the Hall and electron pressure gradient terms and an upwind constrained transport scheme \cite{LondrilloDelZanna2004} with divergence-free reconstruction of magnetic field \cite{Balsara2009}. 

Vlasiator's explicit solvers decompose the solution of the Vlasov equation
\begin{equation}
  \frac{\partial f_s}{\partial t} + \vec{v} \cdot \frac{\partial f_s}{\partial \vec{x}} + \frac{q_s}{m_s}
  \left( \vec{E} + \vec{v} \times \vec{B} \right) \cdot
  \frac{\partial f_s}{\partial \vec{v}}
      = 0
      \label{eq:vlasov}
\end{equation}
into six Cartesian shear operations per time step $\Delta t$. These are constructed as semi-Lagrangian updates \cite{Palmroth2018LRCA}, utilizing a stencil the size of which is dictated by the selected polynomial fitting. Propagation in position space ($\textbf x$) is referred to as \emph{translation}. The Lorentz force effect (first and third terms of equation (\ref{eq:vlasov}), evolution in velocity space ($\textbf v$), termed \emph{acceleration}) is deconstructed into three Cartesian direction aligned shear transformations using the SLICE-3D \cite{slice3d} algorithm. These two solver steps are staggered in time as per Strang splitting \cite{Strang1968}. In this investigation, we focus on the translation solver (utilizing usually a stencil of 2 cells), and for details on other solvers, we refer to \cite{Palmroth2018LRCA}. The 3D spatial mesh in Vlasiator is managed by the DCCRG grid library \cite{dccrgpaper}, and supports dynamic cell-based octree adaptive mesh refinement (AMR) \cite{Ganse2023PoP,Kotipalo2024GMD}.

A notable aspect of Vlasiator's grid implementation is that of a \emph{sparse velocity space}, where the velocity space which is stored and propagated evolves over time \cite{vonAlfthan2014JASTP,Palmroth2018LRCA}, with 64-element \emph{blocks} of phase-space cells added to or subtracted from the velocity space where necessary. The block size can be adjusted for suitable architectures \cite{battarbee2024portinggridbased3d3vhybridvlasov}. This results in the computational complexity of each spatial cell changing as a function of time, position, and surrounding dynamics, up to several orders of magnitude. As a result of mesh refinement, the 2-cell stencil region required for semi-Lagrangian updates may span up to 8 cells (when reaching from a lower refinement region into a higher refinement region).

The Vlasiator code is openly available under the GNU GPL-2 license \cite{Vlasiator}. Performance metrics can be gathered, for example, using the PHIPROF library \cite{phiprof} or via the TAU analysis tools \cite{tau}.

\subsection{Translation solver}

Spatial advection with the translation solver of Vlasiator consists of several steps. Solving the first two terms of the Vlasov equation (\ref{eq:vlasov}) is decomposed into three Cartesian shear motions. Vlasiator performs these in fixed order, historically defined as $Z \rightarrow X \rightarrow Y$. Thus, each task must complete the $Z$-direction translation before being able to proceed to the $X$-direction, which in turn must complete before solving the $Z$-direction. The semi-Lagrangian re-mapping used in Vlasiator solves the evolved state of the distribution function by fitting a polynomial function to phase-space sampling points, advecting the polynomial according to the correct speed (in Cartesian translation operations, this is the velocity $v_x$, $v_y$, or $v_z$ associated with that position in phase space), and integrating the resultant contribution of the advected distribution function back into the discretized grid, storing volume averages of the distribution function $f(\textbf x, \textbf v)$. For the piece-wise parabolic fitting utilized in translation, Vlasiator requires a stencil of 2 cells around each active cell in both directions. Results are written into the active cell itself as well as into face-neighbouring cells in the directions of propagation. The semi-Lagrangian method is not constrained by a Courant-Friedrichs-Lewy condition \cite{courant1928partiellen}. Instead, the limiting factor is having write access to the target cells required by the advected distance. For simplicity and memory management, Vlasiator limits the translation time step so that a single face neighbour cell is sufficient.

\begin{figure}
\includegraphics[width=0.88\textwidth]{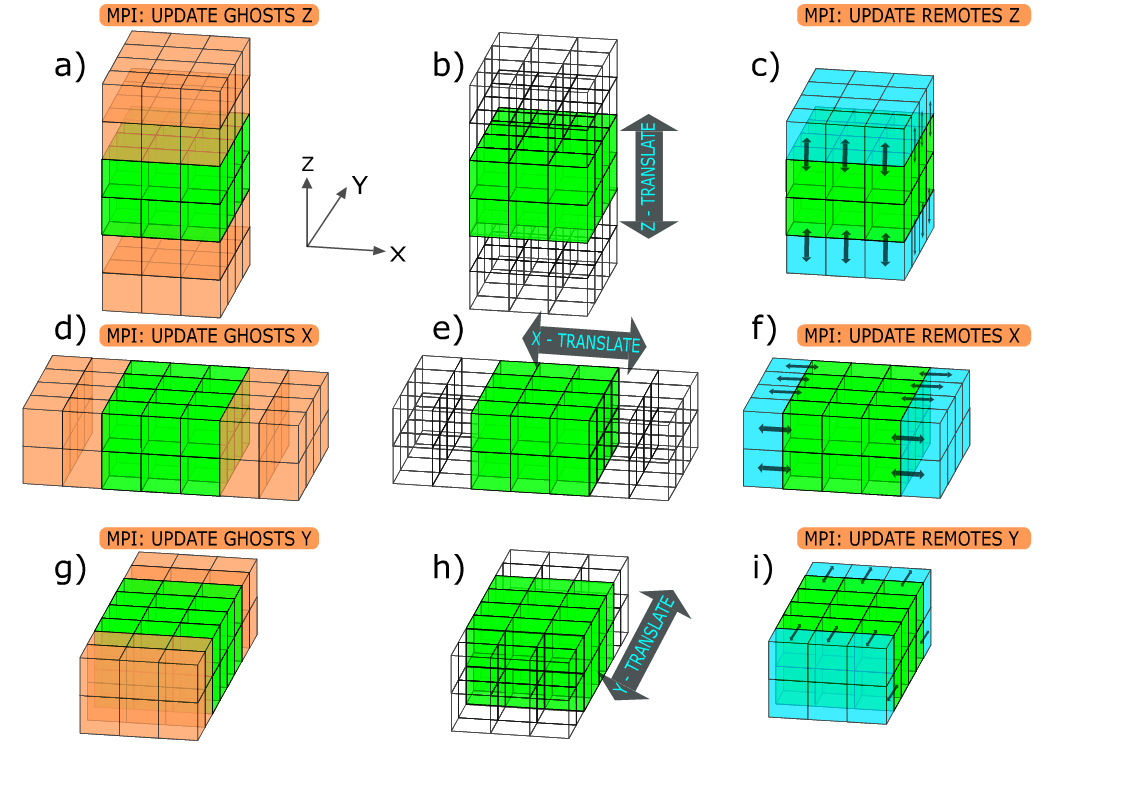}
\caption{Classical spatial Vlasov translation in Vlasiator. For each Cartesian direction, first, ghost domains (orange) are updated over MPI (panels a,d,g), then local (green) and ghost (wireframe) data is used to calculate advection (panels b,e,h), and finally remote neighbour contributions (cyan) are communicated over MPI (panels c,f,i).} \label{fig:classical_translation}
\end{figure}

After each spatial translation direction has been calculated and resultant phase-space density has been assigned to both local and face-neighbouring cells, the phase-space density which has flown across MPI task domain boundaries must be communicated. Thus, solving each Cartesian translation direction consists of three operations:
\begin{enumerate}
    \item Update stencil ghost data from MPI neighbours
    \item Calculate spatial advection with semi-Lagrangian algorithm
    \item Communicate sends and receives of contributions across MPI task boundaries (remote neighbour contribution)
\end{enumerate}
This (classical) approach of the Vlasiator translation solver is exemplified in Fig.~\ref{fig:classical_translation}. The domain in the figure has been simplified significantly, limiting it to a rectangular domain in the absence of mesh refinement. In a real-world situation, a single MPI task may contain cells on several different refinement levels and domains are not limited to rectilinear shapes.

In simulations where the DCCRG mesh is cell-based octree-refined, each spatial cell may have neighbours which are on a larger or smaller refinement level. When a neighbour within the translation stencil is on a higher refinement level, it by extension has similar-sized siblings also within the same stencil. Thus, a single cell participating in translation must solve 4 translation calculations, one for each \emph{pencil} constructed along the AMR mesh. For a detailed description of pencils in the translation solver, see \cite{Ganse2023PoP}.

\subsection{Load balancing Vlasiator}\label{sec:lb}

On a distributed computing system, the DCCRG grid library utilizes Zoltan \cite{zoltan} to partition the spatial mesh to MPI tasks. Each spatial cell is always confined to a single task, so that all data required for acceleration in velocity space is local to that task. The weight assigned to each cell for load balancing purposes is a function of the complexity of the local velocity space and, for example, how many spatial remapping operations (pencils, see \cite{Ganse2023PoP}) the cell participates in. 

\begin{figure}
\includegraphics[width=0.49\textwidth]{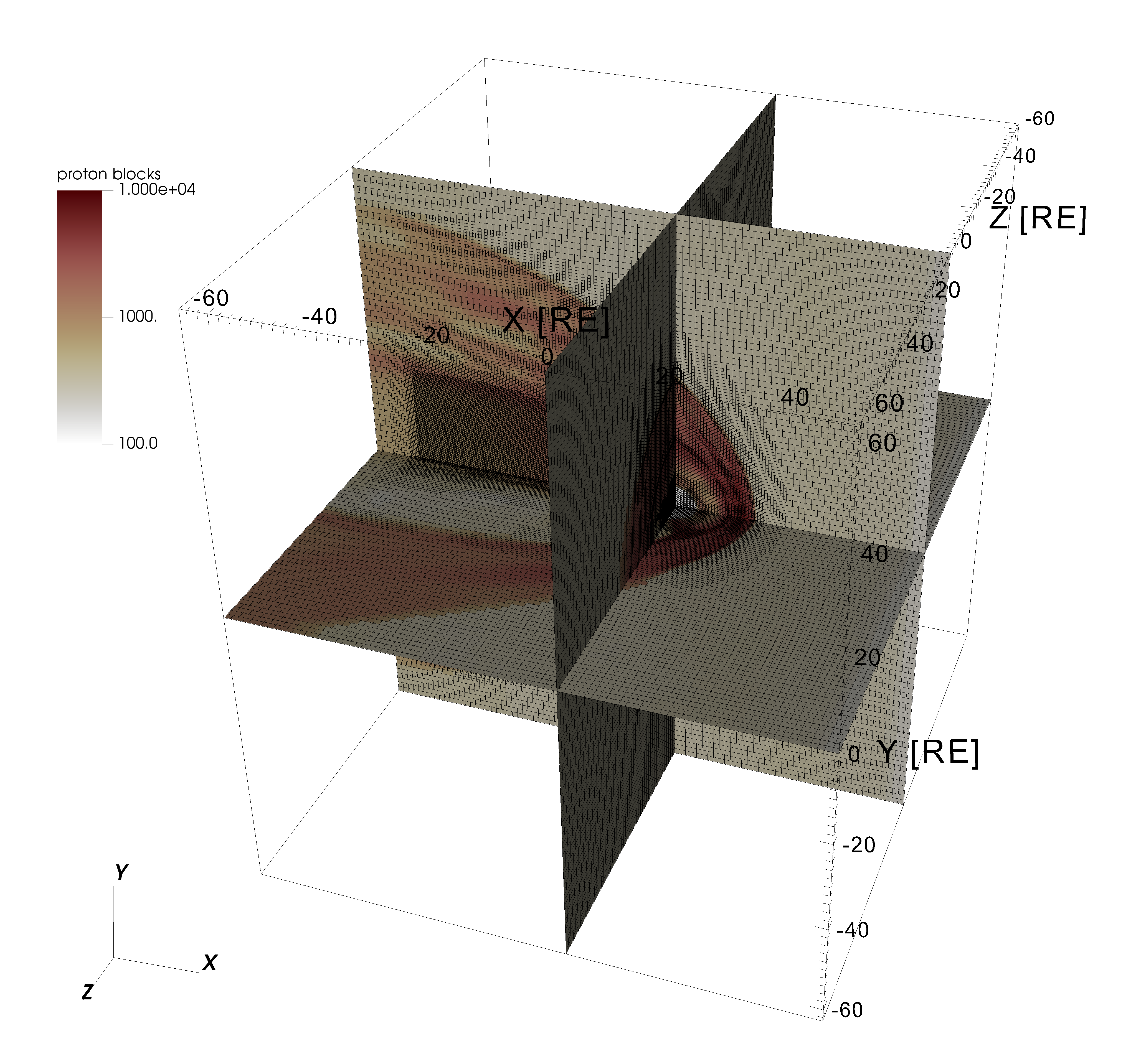}
\includegraphics[width=0.49\textwidth]{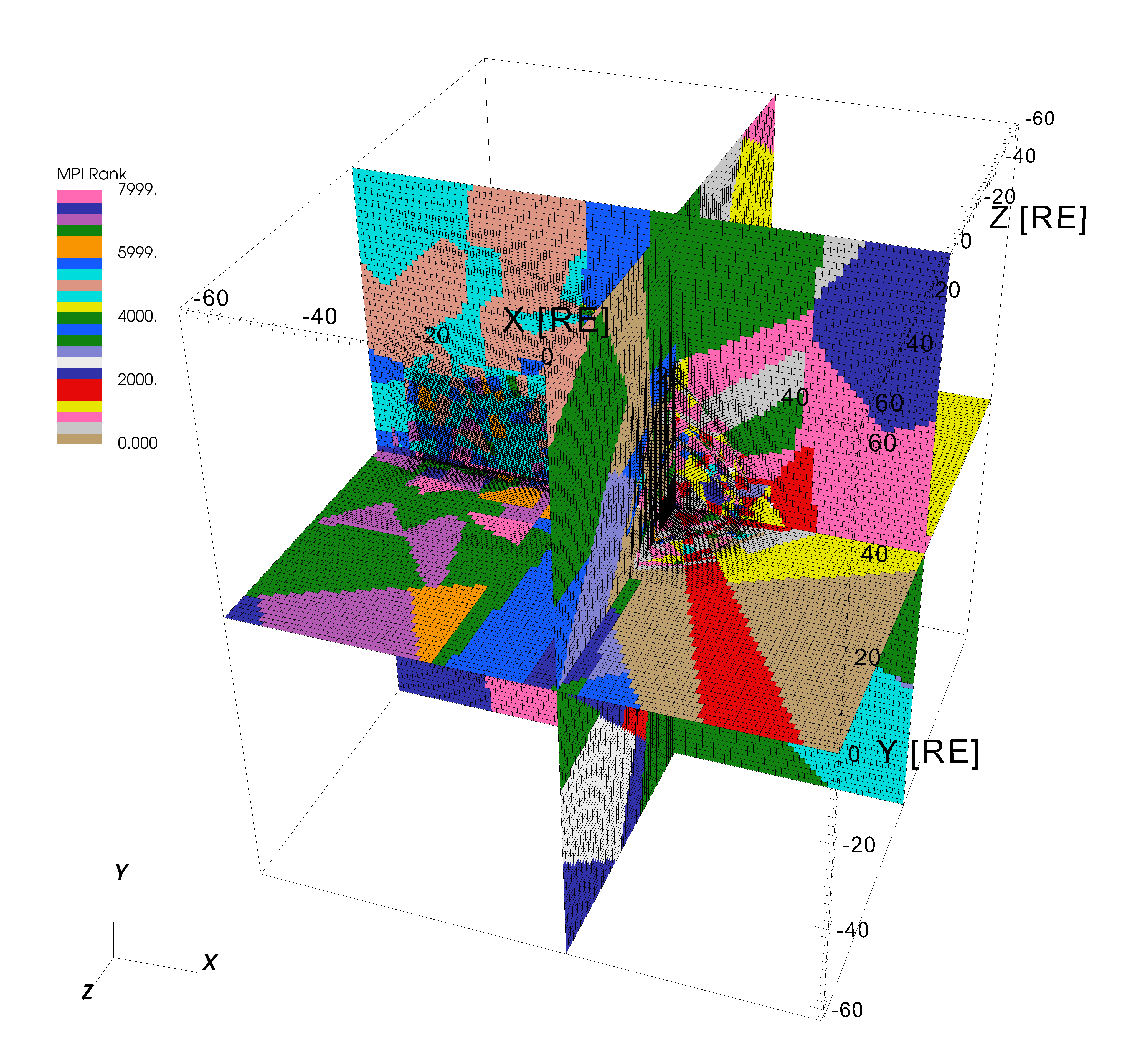}
\caption{
A Vlasiator global 6D simulation at \mbox{$t=800~\mathrm{s}$} with northward magnetic field solar wind driving. Left: The simulation complexity indicated as the count of \emph{blocks} (64-element units of phase-space density), with the dynamically adapting spatial mesh overlaid. Right: MPI task distribution of the simulation domain over 8000 tasks using the Recursive Inertial Bisection (RIB) algorithm \cite{zoltan}.} \label{fig:FIC}
\end{figure}

A sample Vlasiator global simulation, modelling the Earth's magnetic domain under positive $B_z$ solar wind driving conditions, is shown in Fig.~\ref{fig:FIC}. The simulation in question has a base level spatial grid of $90 \times 100 \times 100$ cells at a resolution of $8000~\mathrm{km}$, and 3 levels of dynamic mesh refinement, going up to $1000~\mathrm{km}$. The simulation at the depicted time of 800 seconds consists of 766~474~level~0 cells, 764~754~level~1 cells, 3~057~749~level~2 cells, and 5~199~064~level~3 cells, with roughly $10^{12}$ phase-space cells in total. The simulation was run on 8000 MPI tasks with 16 OpenMP threads each. The solar wind driving has a temperature of $5\cdot 10^5~\mathrm{K}$, density $n_\mathrm{p}=1\cdot 10^6~\mathrm{m}^{-3}$, and flow velocity $V_x = -750~\mathrm{km~s}^{-1}$, as well as an interplanetary magnetic field of $B_z = 5~\mathrm{nT}$. The velocity resolution is set to $40~\mathrm{km~s}^{-1}$. Fig.~\ref{fig:FIC} showcases the refined mesh, the varying number of proton blocks per spatial cell, and the MPI decomposition using the Zoltan Recursive Inertial Bisection (RIB) algorithm \cite{zoltan}.

Due to varying velocity space complexity, non-uniform spatial domains, and mesh refinement, the computational cost of translating data in each Cartesian direction can vary between directions and MPI tasks. Similarly, the amount of data required by the stencil MPI ghost communications can vary.
The interim MPI update requirements of the classical Vlasiator translation solver leads to frequent synchronization requirements and significant wait times between internal steps of the solver, resulting in inefficient use of resources.

\section{Coalescing MPI: the Ghost Translation solver}

We now describe how the Vlasiator translation solver can be re-designed to communicate a large region of ghost data in one fell swoop, then proceeding to solve the semi-Lagrangian translation update for some ghost cells in addition to local cells. This approach, termed \emph{Ghost Translation}, communicates and computes more data than the classical method, but generates only one MPI synchronization point. The overhead of communication and computation is a function of MPI task surface area and mesh complexity of ghost regions.

\subsection{Selecting active and source cells}

In translation, a semi-Lagrangian re-mapping procedure is triggered for each active spatial cell, and for each velocity space element. For the given velocity space element, stencil data is read from spatial neighbours, a polynomial mapping is performed, and integrated phase-space density is stored into the target buffer. Thus, in the classical translation solver, active cells are always local to the MPI task, and can be non-boundary simulation cells or adjacent to them. This adjacency requirement is so that constant boundary cells can cause plasma to flow into actual simulation cells. Inactive cells in translation would be e.g. boundary cells beyond the first layer, but they would still participate in forming translation stencils and providing source data for the polynomial fitting.

\begin{figure}
\includegraphics[width=0.79\textwidth]{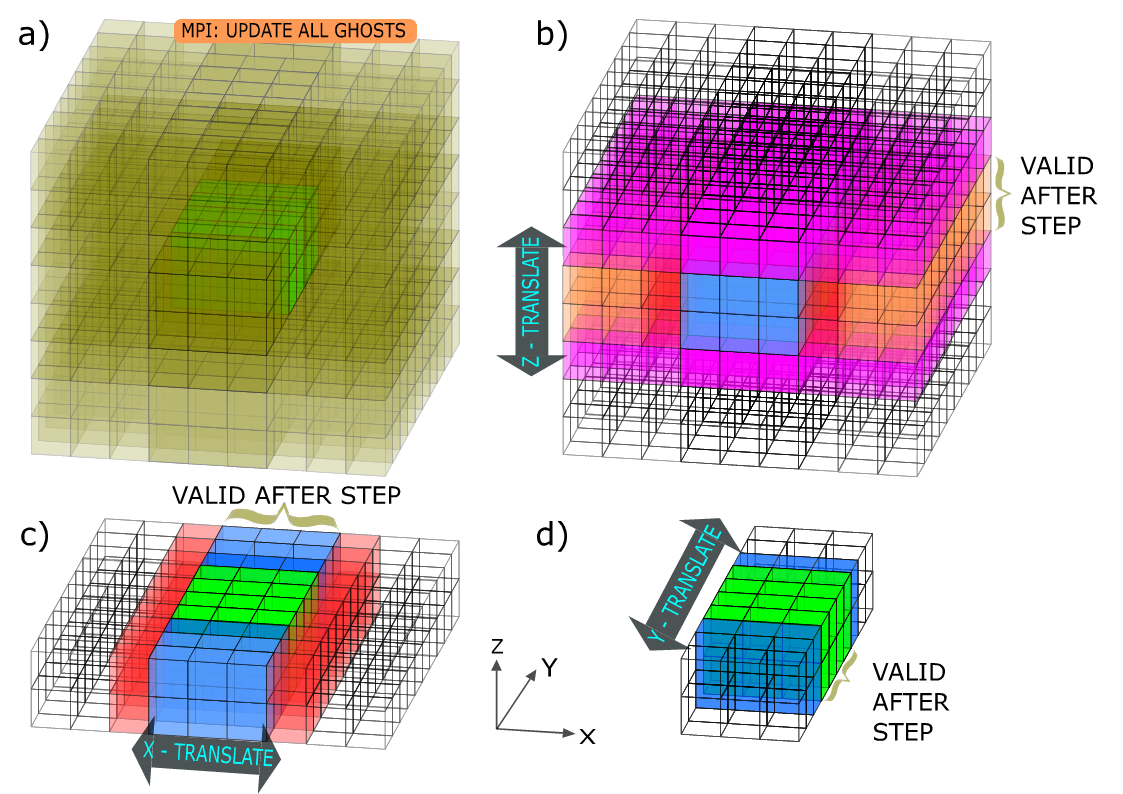}
\caption{New Ghost Translation method steps: a) The translucent olive ghost cells around the green local domain (partially obscured in the centre) are updated via MPI communication. b) $Z$-directional translation is evaluated for all $Z$-active cells (coloured, ghost and local). c) $X$-directional translation is evaluated for all $X$-active cells (coloured, ghost and local). d) $Y$-directional translation is evaluated for all $Y$-active cells (coloured, ghost and local). The result is correct data in the local domain.} \label{fig:ghost_translation}
\end{figure}

In order to perform translation in the 3 Cartesian directions (in order $Z \rightarrow X \rightarrow Y$) and thereafter have valid local data, we must now mark some ghost cells as active cells so that they may act as plasma inflow sources to the local domain. To evaluate this, we proceed in reverse order of Cartesian directions:
\begin{itemize}
    \item In the final step ($Y$-direction), we designate all local cells and all their $Y$-directional face neighbours as $Y$-active cells.
    \item For each $Y$-active cell, we evaluate a 2-cell wide translation stencil in the $Y$-direction. These cells are tagged as $Y$-sources.
    \item Stepping back, in the $X$-direction: we designate all $Y$-source cells and their $X$-directional face neighbours as $X$-active cells.
    \item For each $X$-active cell, we evaluate a 2-cell wide translation stencil in the $X$-direction. These cells are tagged as $X$-sources.
    \item Reaching the first step, now evaluating the $Z$-direction: we designate all $X$-source cells and their $Z$-directional face neighbours as $Z$-active cells.
    \item Finally, for each $Z$-active cell, we evaluate a 2-cell wide translation stencil in the $Z$-direction. These cells are tagged as $Z$-sources.
\end{itemize}
Cells marked as sources must be updated before performing Ghost Translation, but each Cartesian direction can proceed in the $Z \rightarrow X \rightarrow Y$ order regardless of other MPI task states. This process is exemplified, for a simple non-refined rectangular MPI domain, in Fig.~\ref{fig:ghost_translation}.

\subsection{Limiting active cells}

Ghost Translation allows an additional trade-off between computation, communication, and numerical accuracy. When selecting which cells to compute (active cells), one can either require all cells required for later stencils to be computed, including both themselves and their face neighbours (the full approach, dubbed GT3). Alternatively, one can require as little as one layer of cells around the local domain to be computed (the minimal approach, dubbed GT1). By extension, an intermediate version (GT2) is also possible. 
If data in stencils are not updated in these computation steps, they will retain their MPI updated values.


\subsection{Estimated overhead}\label{sec:overhead}

\begin{figure}
\includegraphics[width=0.9\textwidth]{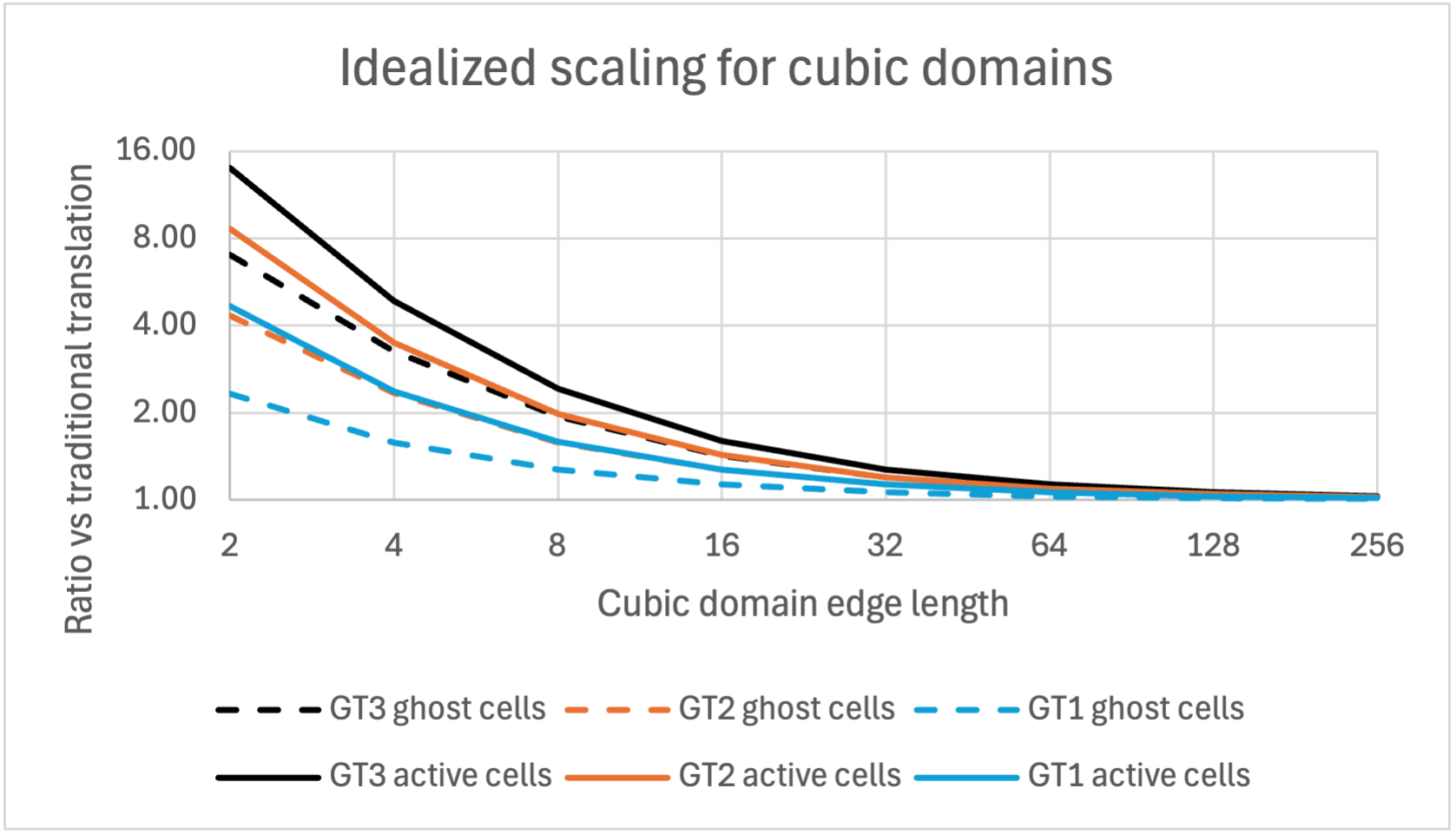}
\caption{Ideal scaling for Ghost Translation assuming uniform cubic domains. A full ghost translation stencil (GT3, black) results in excessive overhead in both MPI ghost cell update requirements (dashed line) as well as required computations (solid line). An intermediate stencil (GT2, orange) has less overhead and the minimal (GT1, blue) even smaller overhead. For all approaches, the overhead decreases as a function of the inner volume of the domain, being no more than 27\% for edge lengths of 32 cells.} \label{fig:ratios}
\end{figure}

The increases in MPI communication and computation will vary with mesh refinement and shape of each MPI domain. Additionally, MPI tasks touching simulation boundaries will not require extra Ghost Translation layers at those edges, resulting in smaller overheads. A rough estimate can, however, be achieved by assuming a non-refined grid consisting of only cubic domains with periodic boundary conditions. Relative overheads for ideal cubic domains of increasing sizes are 
graphed in Fig.~\ref{fig:ratios}.

\section{Performance results}

To evaluate the real-world performance of Ghost Translation, we ran parameter studies of both artificial scaling tests and production-scale global magnetospheric simulations. We performed these tests utilizing identical simulation setups for classical Vlasiator translation (base), Ghost Translation with minimal domains (GT1) and Ghost Translation with full domains (GT3). All tests were performed on the Mahti supercomputer at CSC in Kajaani, Finland. A single Mahti node contains two AMD Rome 7H12 CPUs with 64 cores each. The CPUs are based on the AMD Zen 2 architecture, supporting the AVX2 vector instruction set, and run at 2.6 GHz base frequency (max boost up to 3.3 GHz). 


\subsection{Weak scaling}

We now demonstrate weak scaling by running a 6D simulation of a plasma pulse traveling into a region with spatial mesh refinement in the central area. The background proton plasma temperature was set to $10^5~\mathrm{K}$, density to $n_\mathrm{p,0}=10^6~\mathrm{m}^{-3}$, and flow velocity to $V_x = 100~\mathrm{km s}^{-1}$. From one end of the tube, new hot plasma flows in with temperature $2\cdot 10^6~\mathrm{K}$, density $n_\mathrm{p}=2\cdot 10^6~\mathrm{m}^{-3}$, and flow velocity $V_x = 200~\mathrm{km~s}^{-1}$. Throughout the box, an initial constant magnetic field of $(1,1,1)~\mathrm{nT}$ is set. The spatial resolution at level 0 is set to $2\cdot10^4~\mathrm{km}$, and the velocity resolution is $30~\mathrm{km~s}^{-1}$. Acceleration, translation and field solvers are all active throughout the simulation. We evaluate performance by averaging over simulation steps 109--134, in order to have a more realistic simulation state with plasma having evolved in the domain.

For this weak scaling test, the base grid (at refinement level zero) was $20\times 32 \times 4$ cells. With refinement up to three times, this resulted in 1~840~level~0 cells, 3~072~level~1 cells, 9~216~level~2 cells, and 98~304~level~3 cells for a single-node run. Other simulation parameters are kept constant, but the $Z$-directional extent and spatial cell counts are multiplied by the node count, resulting in constant computational effort per task. We note, however, that this artificial test is an idealized scenario and does not suffer of the same imbalances as production-scale magnetospheric simulations do. Runs were performed on 1, 2, 4, 16, 64, and 128 nodes. Due to constantly evolving velocity meshes, the count of phase-space cells per spatial cell ranged from 13~120 to 100~224 elements per cell with an average evolving from 15~168 to 17~280 elements throughout the simulation. Load balance was performed with the RIB method of Zoltan \cite{zoltan} every 10 steps of simulation.

\begin{figure}
\includegraphics[width=0.49\textwidth]{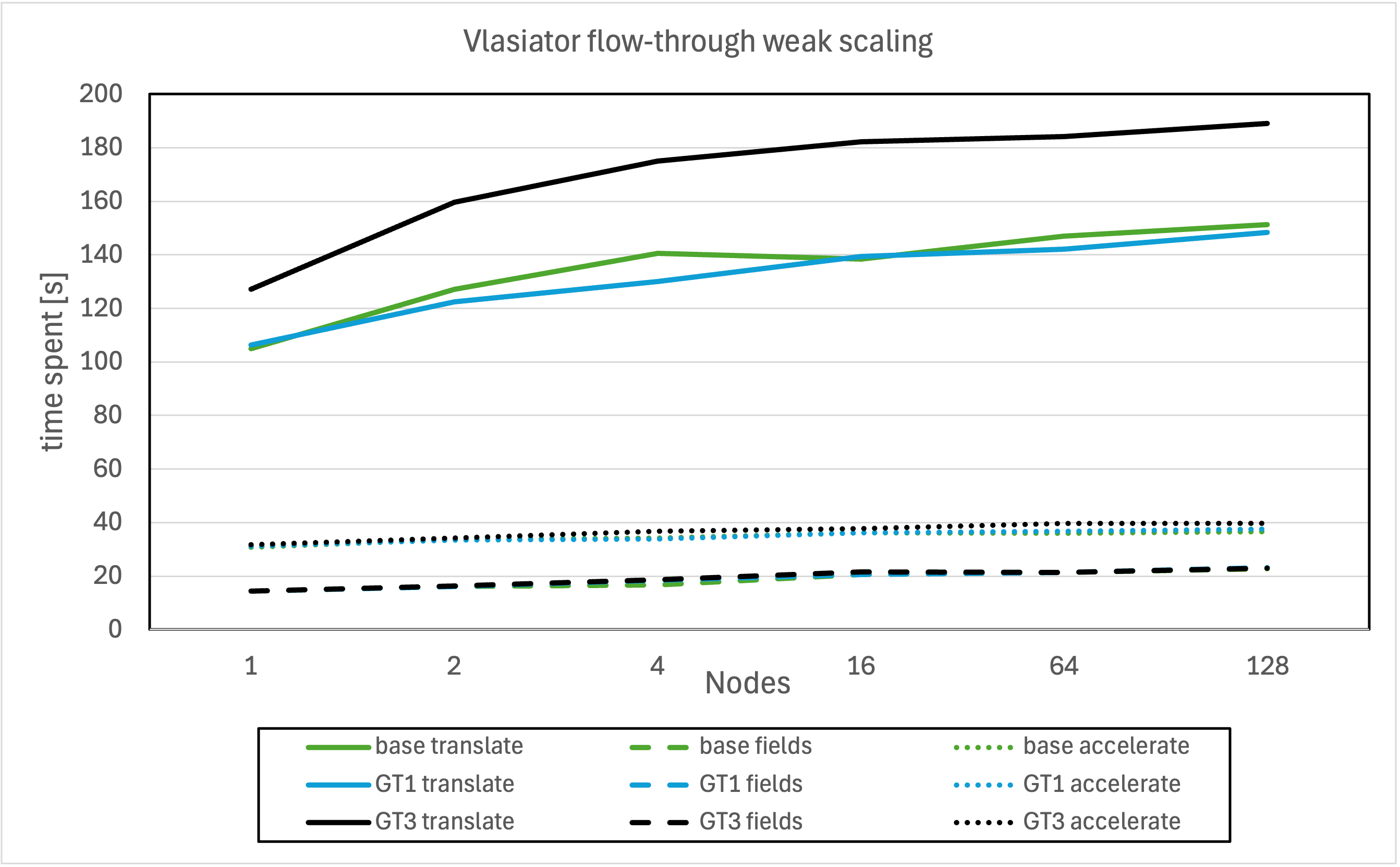}
\includegraphics[width=0.49\textwidth]{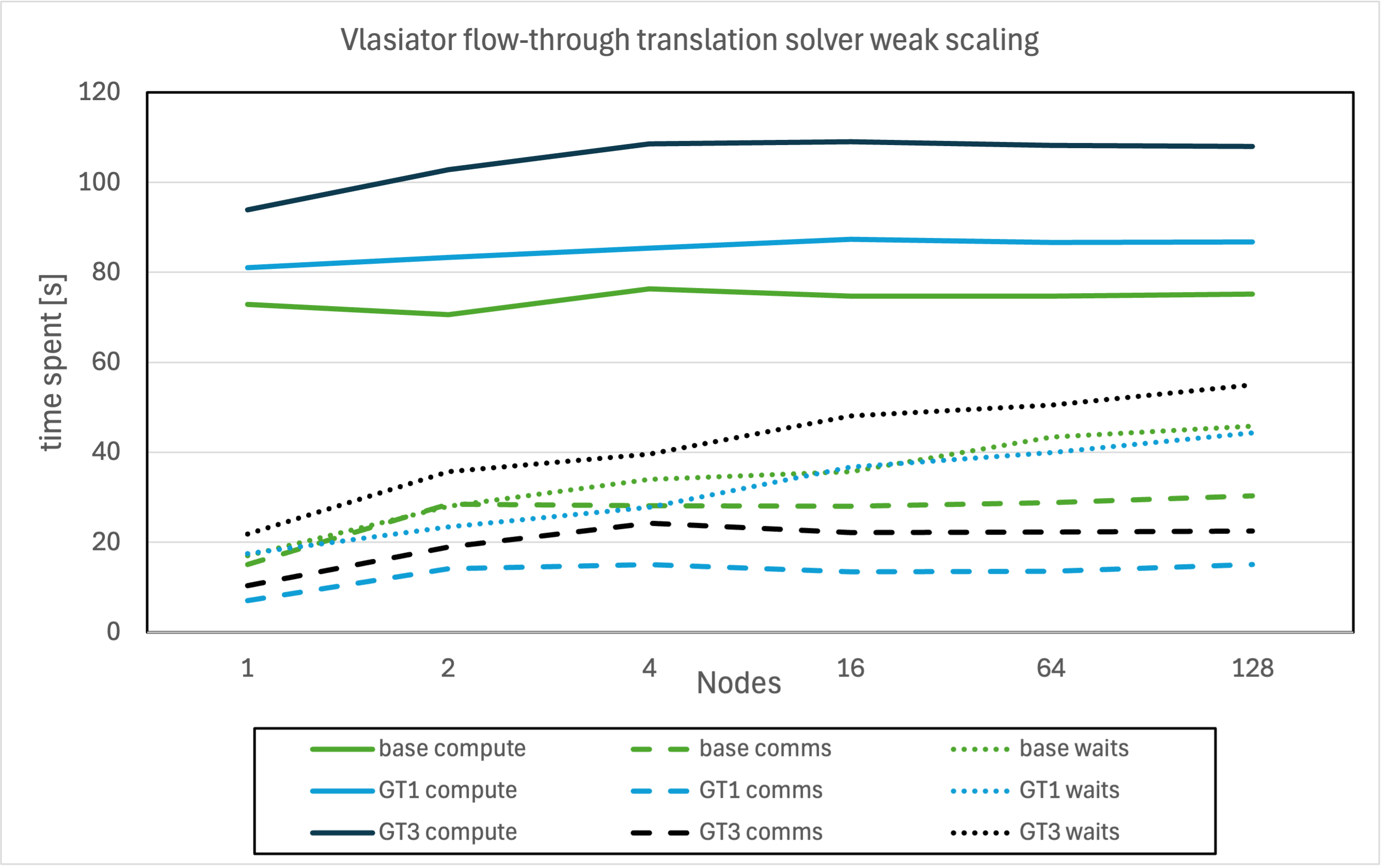}
\caption{
Left: Weak scaling results for the Vlasiator solvers 
for a flow-through plasma simulation with 1\ldots128 nodes, 4 tasks per node, and 3 levels of mesh refinement. 
Right: Decomposition of time spent in the translation solver into computation, communication, and waiting. Results are shown for classical translation (base, green), full stencil Ghost Translation (GT3, black), and minimal stencil Ghost Translation (GT1, blue).
} \label{fig:weakscaling}
\end{figure}

In Fig.~\ref{fig:weakscaling}, the left-hand panel shows weak scaling performance of Vlasiator.
The acceleration solver (dots) acts wholly on local data and should scale ideally, but in larger simulations, load balance may become imperfect. The field solver (dashes) includes MPI ghost communication, but scales near-ideally nevertheless. The translation solver performance (lines) is shown for the classical translation (green) as well as the GT1 (blue) and GT3 (black) methods.

In the right-hand panel of Fig.~\ref{fig:weakscaling}, we show the scaling for the time division within the translation solver into computation, communication, and waiting for other processes. Time spent in computation (lines) or communication (dashes) shows ideal scaling when evaluated at 4 or more nodes. Interestingly, although the actual amount of communication in Ghost Translation is greater than in the regular version, both GT versions outperform the base version due to being able to do a single communication call. The worst scaling is found in time spent in waits (dots) -- this is due to imbalance of the workload. GT1 (blue) slightly outperforms the base version (green) in scaling for waiting time, but GT3 (black) lags behind in the current tests.

\subsection{Performance at scale}

To evaluate the real-world performance of the Ghost Translation approach, we gathered performance metrics over 50 seconds of simulation, restarting from a saved checkpoint state of the global magnetospheric simulation described in section \ref{sec:lb} with the classical, GT1, and GT3 approaches. The simulation used 500 compute nodes, 8000 MPI tasks and 16 OpenMP threads each. We perform the same test with three Zoltan load balance approaches: Recursive Coordinate Bisection (RCB), RCB constrained to rectilinear domains, and Recursive Inertial Bisection (RIB) \cite{zoltan}.

\begin{table}
\caption{Results at scale: 50 steps of a global 6D magnetospheric simulation. Columns are 3 different translation approaches each with 3 different load balance algorithms. Rows indicate time spent in computation, communication, interim and final waits (min. values in bold), as well as Ghost Translation cell count ratios and memory statistics.}\label{tab:results}
\begin{tabular}{|l||l|l|l||l|l|l||l|l|l|}
\hline
 & base & GT1 & GT3 & base & GT1 & GT3 & base & GT1 & GT3 \\
\hline
Load balance algorithm & \multicolumn{3}{|c||}{RCB} & \multicolumn{3}{|c||}{RCB rectilinear} & \multicolumn{3}{|c|}{RIB}\\
\hline\hline
Compute [s] & \textbf{50} & 80 & 141 & \textbf{49} & 76 & 134 & \textbf{51} & 86 & 163 \\
\hline
Communication [s] & 117 & \textbf{95} & 176 & 111 & \textbf{89} & 164 & 129 & \textbf{95} & 201 \\
\hline
Total waits [s] & \textbf{292} & 294 & 467 & 303 & \textbf{276} & 442 & \textbf{231} & 244 & 419 \\
\hline
Total translate [s] & \textbf{459} & 473 & 790 & 464 & \textbf{445} & 747 & \textbf{412} & 430 & 792 \\
\hline\hline
Avg active cells ratio & 1 & 1.99 & 5.60 & 1 & 1.91 & 5.40 & 1 & 2.18 & 6.49 \\
\hline
Avg source cells ratio & -- & 1.99 & 3.96 & -- & 1.92 & 3.81 & -- & 2.18 & 4.55 \\
\hline\hline
Avg mem resident [GiB] & 70.3 & 82.5 & 133.1 & 68.6 & 81.0 & 128.7 & 75.2 & 87.4 & 160.8 \\
\hline
Avg mem high mark [GiB] & 84.3 & 96.9 & 146.5 & 82.4 & 95.7 & 142.3 & 91.6 & 102.6 & 173.0 \\
\hline
\end{tabular}
\end{table}

The performance statistics shown in Tab.~\ref{tab:results} indicate that in its current form, Ghost Translation with minimal stencils is able to roughly match (or in the case of rectilinear RCB domains, exceed) the classical approach, but the computational overhead from full stencils leads to excess time spent in the range of 60--90\%. This is explained by the ratios for active and source cells being quite high \mbox{($\sim2 \ldots \sim 6.5$).} This is much larger than the source ratios of 1.20 (GT1) and 1.67 (GT3) or the active ratios of 1.42 (GT1) and 1.98 (GT3) which regular cubic domains at the mean of 1223 cells per task would have (see section \ref{sec:overhead}), indicating that mesh refinement of Ghost Translated regions has a significant impact on performance. The increase in included ghost domains, in particular near mesh refinement, leads also to increases in memory use. Resident memory increased by up to 114\% (GT3) or 18\% (GT1), with high water mark memory measured at up to 89\% (GT3) or 16\% (GT1) more.


\section{Discussion and conclusions}

It is not surprising that the Ghost Translation methodology is not an outright performance improvement. It is a trade-off, communicating more cells in one go instead of doing direction-by-direction updates. It is also a trade-off performing extra computation instead of communicating updates. Thus, the outright performance change is a function of MPI network performance and local computing power. As such, it may become quite useful when applied to accelerated hardware architectures \cite{battarbee2024portinggridbased3d3vhybridvlasov}, where local computation may become more beneficial. Additionally, though Ghost Translation was envisioned in order to tackle load balancing challenges in the classical approach, the tested load balancing algorithms may not be optimal for Ghost Translation. Instead, a method which minimizes task surface area (and thus Ghost Translation overhead) such as a graph-based approach might be preferable. In particular, the new load balancing approach would need to account for refinement of Ghost Translated regions accruing a performance penalty due to more required ghost communication and computation. This is, however, left to a further study.

Another, potentially more significant effect is how Ghost Translation can act in tandem with other algorithmic improvements. If a simulation domain contains strong magnetic fields such as the dipole field of Earth, the maximum allowed time step may depend strongly on the spatial location of the simulation cell. As such, it may become beneficial to implement local time stepping, using different values of $\Delta t$ in different spatial domains. In order to evaluate the spatial advection in that region, the $\Delta t$ used for each class of cells must be used both for local translation and also for remote neighbour contribution -- necessitating a method such as Ghost Translation at boundaries between different $\Delta t$ classes.

In addition, as any imbalance in Ghost Translation is at the end of that code region, a task-based approach could perform additional computations whilst waiting for other tasks to complete their Ghost Translation work. In classical translation, this is not as feasible due to the waiting period being split into several small waits per Cartesian direction.



\begin{credits}
\subsubsection{\ackname} 
This work was made possible through the EuroHPC ``Plasma-PEPSC'' Centre of Excellence (Grant number 4100455) and the Research Council of Finland matching funding (grant number 359806). The Research Council of Finland is also acknowledged for funding through grants 
335554, 
339756, 
339327, 
312351 and 
336805, 
which have supported Vlasiator development and science.

Vlasiator \cite{Vlasiator} is open-source under the GNU GPL-2 license and hosted at GitHub (\url{https://github.com/fmihpc/vlasiator}). The datasets used and presented in this work requires multiple terabytes of storage and dedicated infrastructure for its handling and analysis. It can be shared upon reasonable request to the authors.

\subsubsection{\discintname}
The authors have no competing interests to declare that are
relevant to the content of this article. 
\end{credits}

%
%
%
%

\end{document}